\documentclass[lnbip]{svmultln}
\usepackage{framed,enumitem}
\usepackage{graphicx}
\usepackage{makeidx}
\usepackage{listings}
\usepackage{xcolor}

\lstdefinestyle{python}{
    language=Python,
    basicstyle=\ttfamily\small,
    commentstyle=\color{green!40!black},
    keywordstyle=\color{blue},
    numberstyle=\tiny\color{gray},
    numbers=left,
    stepnumber=1,
    frame=single,
    rulecolor=\color{black!30},
    backgroundcolor=\color{gray!5},
    breaklines=true,
    breakatwhitespace=true,
    showspaces=false,
    showstringspaces=false,
    showtabs=false,
    morekeywords={as},
}

\begin{document}

\mainmatter              

\title{Building BESSER: an open-source low-code platform \thanks{This project is supported by the Luxembourg National Research Fund (FNR) PEARL program, grant agreement 16544475}}
\titlerunning{BESSER better software faster}

\author{Iván Alfonso\inst{1} \and
Aaron Conrardy\inst{1} \and
Armen Sulejmani\inst{1} \and
Atefeh Nirumand\inst{1} \and
Fitash Ul Haq\inst{1} \and
Marcos Gomez-Vazquez\inst{1} \and
Jean-Sébastien Sottet\inst{1} \and
Jordi Cabot\inst{1,2}}
\authorrunning{Iván Alfonso et al.}

\institute{Luxembourg Institute of Science and Technology, Luxembourg \\
\email{{ivan.alfonso,aaron.conrardy,armen.sulejmani,atefeh.nirumand,fitash.ulhaq,marcos.gomez,jean-sebastien.sottet,jordi.cabot}@list.lu} \and
University of Luxembourg, Luxembourg}
\maketitle

\newcommand{\nb}[2]{
    \fbox{\bfseries\sffamily\scriptsize#1}
    {\small\textit{#2}\\}
   }
\newcommand\notes[1]{\textcolor{red}{\nb{original notes}{#1}}}
\newcommand\aaron[1]{\textcolor{blue}{\nb{AARON}{#1}}}
\newcommand\FH[1]{\textcolor{red}{\nb{Fitash}{#1}}}
\newcommand\ivan[1]{\textcolor{purple!100}{\nb{Ivan}{#1}}}
\newcommand\marcos[1]{\textcolor{blue!100}{\nb{Marcos}{#1}}}



\newcommand{\OCLtext}{}
\newcommand{\OCLkeyword}[1]{{\OCLtext #1}}
\newcommand{\OCLtextcomment}[1]{ {\OCLtext \emph{#1}} }


\newenvironment{ocl}[1][\linewidth]%
  {\begin{minipage}{#1} \OCLtext }%
  {\end{minipage}}
\newenvironment{ocl-boxed}[1][\linewidth]%
  {\begin{Sbox}\begin{minipage}{#1} \OCLtext }%
  {\end{minipage}\end{Sbox}{\setlength\fboxsep{7pt}\cornersize{0.3}\ovalbox{\TheSbox}}}


\newcommand{\OCLand}{\OCLkeyword{and}}
\newcommand{\OCLattr}{\OCLkeyword{attr}}
\newcommand{\OCLbody}{\OCLkeyword{body}}
\newcommand{\OCLcontext}{\OCLkeyword{\textbf{context}}}
\newcommand{\OCLdef}{\OCLkeyword{def}}
\newcommand{\OCLderive}{\OCLkeyword{derive}}
\newcommand{\OCLelse}{\OCLkeyword{else}}
\newcommand{\OCLendif}{\OCLkeyword{endif}}
\newcommand{\OCLendpackage}{\OCLkeyword{endpackage}}
\newcommand{\OCLif}{\OCLkeyword{if}}
\newcommand{\OCLimplies}{\OCLkeyword{implies}}
\newcommand{\OCLin}{\OCLkeyword{in}}
\newcommand{\OCLinit}{\OCLkeyword{init}}
\newcommand{\OCLinv}{\OCLkeyword{\textbf{inv}}}
\newcommand{\OCLlet}{\OCLkeyword{let}}
\newcommand{\OCLnot}{\OCLkeyword{not}}
\newcommand{\OCLoper}{\OCLkeyword{oper}}
\newcommand{\OCLor}{\OCLkeyword{or}}
\newcommand{\OCLpackage}{\OCLkeyword{package}}
\newcommand{\OCLpost}{\OCLkeyword{post}}
\newcommand{\OCLpre}{\OCLkeyword{pre}}
\newcommand{\OCLself}{\OCLkeyword{self}}
\newcommand{\OCLthen}{\OCLkeyword{then}}
\newcommand{\OCLxor}{\OCLkeyword{xor}}

\newcommand{\OCLarrow}{$-{>}$}

\newcommand{\OCLcomment}[1]{{--}{--} \OCLtextcomment{#1}}


\newcommand{\OCLprecond}[2]{\OCLcontext #1 \\ \OCLpre #2}
\newcommand{\OCLpostcond}[2]{\OCLcontext #1 \\ \OCLpost #2}
\newcommand{\OCLprepostcond}[3]{\OCLcontext #1 \\ \OCLpre #2 \\ \OCLpost #3}
\newcommand{\OCLinvariant}[2]{\OCLcontext #1\OCLinv \\ #2}
\newcommand{\OCLindent}[1]{{\parbox{8pt}{~} \hfill \parbox[b]{\linewidth}{\vspace{2pt}#1} }}
\newcommand{\OCLifthenelse}[3]{\OCLif #1 \OCLthen \\ \OCLindent{#2} \OCLelse \\ \OCLindent{#3}  \OCLendif}
\newcommand{\OCLifthen}[2]{\OCLif #1 \OCLthen \\ \OCLindent{#2} \OCLendif}

\begin{abstract}
Low-code platforms (latest reincarnation of the long tradition of model-driven engineering approaches) have the potential of saving us countless hours of repetitive boilerplate coding tasks. However, as software systems grow in complexity, low-code platforms need to adapt as well. Notably, nowadays this implies adapting to the modeling and generation of smart software. At the same time, if we want to broaden the userbase of this type of tools, we should also be able to provide more open source alternatives that help potential users avoid vendor lock-ins and give them the freedom to explore low-code development approaches (even adapting the tool to better fit their needs).
To fulfil these needs, we are building BESSER, an open source low-code platform for developing (smart) software. BESSER offers various forms (i.e., notations) for system and domain specification (e.g. UML for technical users and chatbots for business users) together with a number of generators. Both types of components can be extended and are open to contributions from the community.
\end{abstract}
\section{Introduction} 
Low-code platforms are designed to accelerate software delivery by minimizing hand-coding efforts \cite{richardson2014new}. Low-code platforms can be regarded as a style of Model-Driven Engineering (MDE) \cite{Cabot20, RuscioKLPTW22} with an emphasis on software development.

Low-code platforms hold significant promise, especially in today's software landscape where complexity is on the rise, and software requirements are becoming increasingly demanding. This includes the growing need to develop/deploy/integrate Artificial Intelligence (AI) components to support and provide advanced features. For instance, the adoption of new user interfaces (such as augmented/virtual reality, chat, and voice interfaces), the implementation of intelligent behavior to classify, predict, or recommend information based on user input, and the need to address emerging security and sustainability concerns. These AI-enhanced systems are commonly referred to as \textit{smart software} \cite{cabot2023lowcode}.

We have witnessed an explosion of low-code tools, mostly targeting Enterprise software \cite{Pinho2023Usability}, with Gartner defining Mendix, OutSystems, Microsoft, ServiceNow, Appian and Salesforce as leaders in this space \cite{Gartner2023}. Nevertheless, all these tools excel at ``classical'' data-intensive web applications while presenting severe limitations for more complex scenarios. An example would be the definition of AI-enhanced interfaces, that, if done, usually only consists of adding endpoints to third party applications (e.g. integration of third party chatbots or Appian's OpenAI plugin), acting rather as an add-on than an actual integral component of the application.
Moreover, they are proprietary ecosystems where modelers usually do not even have access to the code generated from their models as applications are transparently deployed in the vendor's cloud. This is convenient but at the same time generates a clear vendor lock-in. 

Given the increasing complexity of software development, we argue there is a need for powerful open source low-code platforms that facilitate the adoption of low-code approaches while offering an extensible environment enabling companies to adapt and tailor the tool to their specific needs, but also tackling applications beyond Enterprise software, such as AR/VR interfaces or digital twins.

To address these needs, we are developing BESSER, short for BEtter Smart Software FastER, an open-source low-code platform for smart software development. BESSER is available in our project repository \cite{github2023repository}. The platform development is a core element in a 5-year funded project on this same topic, guaranteeing the evolution of the platform over the next years. 

The rest of this paper is as follows: 
in Section \ref{sec:besser}, we introduce BESSER, detailing its architecture, languages and generators. Section \ref{sec:verticals} showcases some practical applications of BESSER through case studies. Section \ref{sec:discussion} delves into the primary insights gained from our experiences, and finally, Section \ref{sec:conclusion} concludes the paper.

\section{Overview of the BESSER platform}
\label{sec:besser}
\begin{figure}
\includegraphics[width=\textwidth]{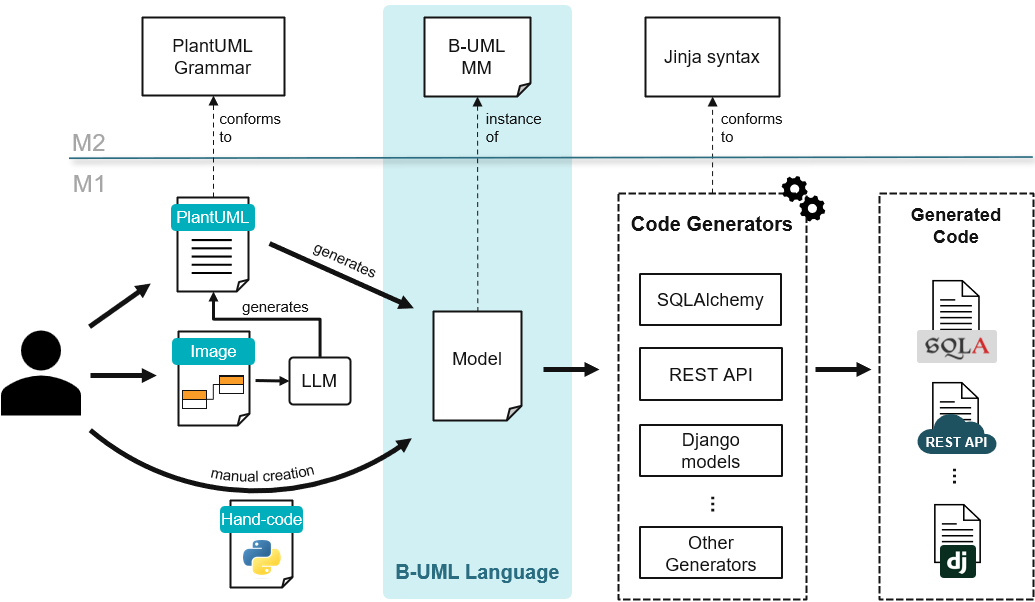}
\caption{BESSER low-code platform architecture} \label{fig:architecture}
\end{figure}

Figure \ref{fig:architecture} illustrates the architecture of the BESSER platform. At the core of this architecture we have B-UML (short for BESSER’s Universal Modeling Language), the foundational language of the BESSER platform used for specifying domain models. B-UML models are then transformed to other models or to software artefacts via model transformations. 

The next sections describe the B-UML language and the code generators provided with it, implemented as model-to-text transformations.  

\subsection{B-UML language}
B-UML, the BESSER Universal Modeling Language, is the base language of the BESSER low-code platform. This language is heavily inspired by UML \cite{specification2017omg} but does not aim to be fully compliant with it. Instead, the goal is to have the freedom to integrate other types of (meta)models while benefitting from parts of the UML that we find interesting for our approach.
It is up to the designer to decide which of these sublanguages to 'activate' for a given project based on their specific modeling requirements.
    
\begin{figure}
\includegraphics[width=\textwidth]{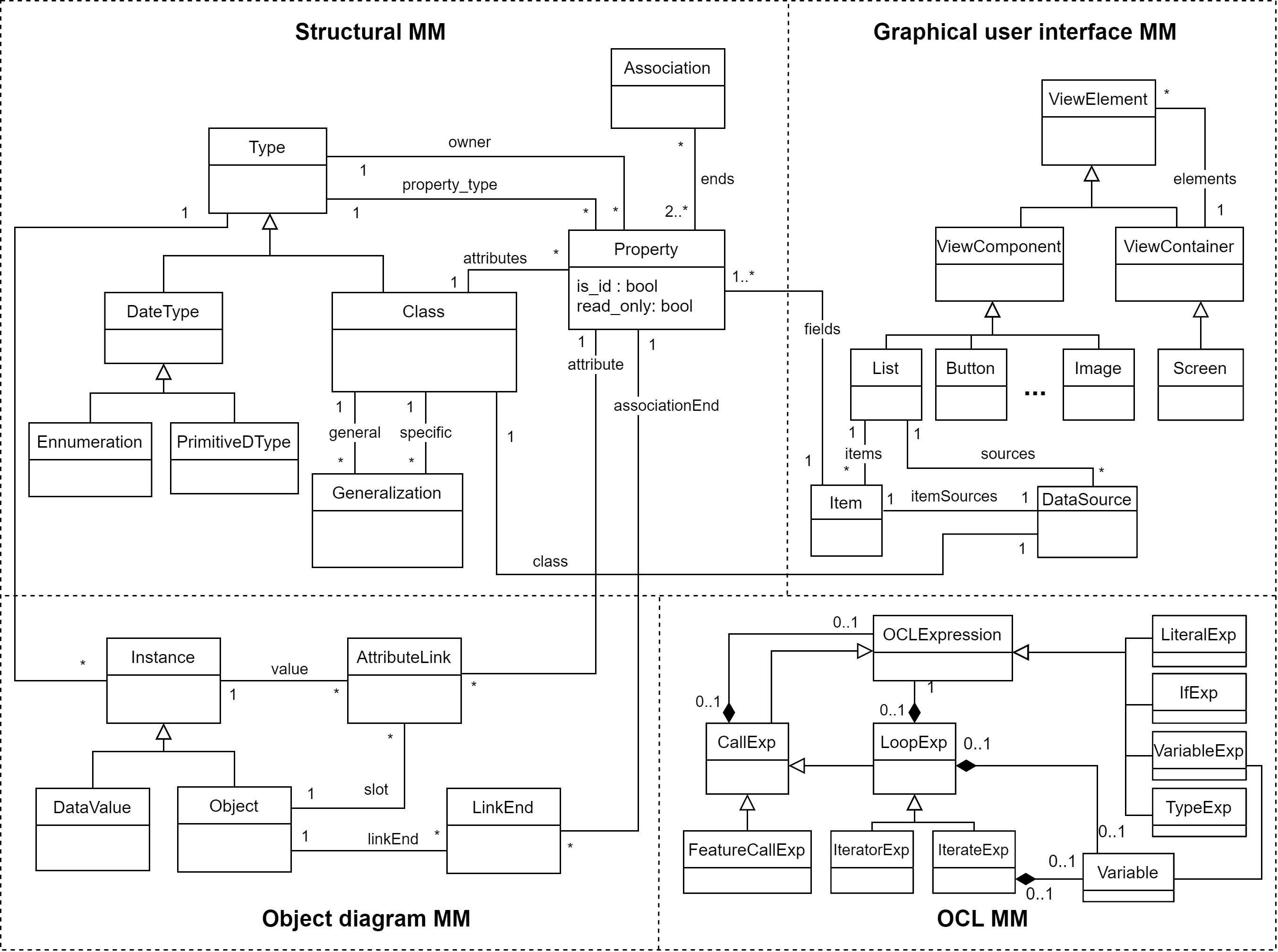}
\caption{Excerpt of the B-UML language metamodel} \label{fig:metamodel}
\end{figure}

Figure \ref{fig:metamodel} depicts an excerpt of the B-UML language divided into submodules. For the sake of brevity and space, this metamodel omits various concepts and relationships, focusing on the most relevant ones. The complete metamodel is available in the project repository \cite{github2023repository}.
Note that these are the submodules or sublanguages currently defined in BUML. However, we have a list of additional sublanguages outlined in our immediate roadmap (Section \ref{sec:roadmap}) that we plan to design and develop to further support BESSER's objectives (e.g., AI component modeling).

\textbf{Structural metamodel:} it enables the specification of a domain model using the typical concepts of a class diagram. Elements such as \textit{Classes}, \textit{Properties}, \textit{Associations}, and \textit{Generalizations} can be instantiated to define the static structure of a system. While this metamodel is rooted in the UML specification, certain modifications and additions have been implemented to provide additional modeling capabilities. For instance, the \textit{is\_id} attribue has been introduced in the Property class to specify whether a property serves as an identifier for the instances of that class, a common need in many code generation scenarios.

\textbf{Object diagram metamodel}: while the structural metamodel centers on classes and their static structures, the object diagram metamodel enables the representation of how these classes are instantiated into objects and interact with each other. The \textit{Object} class in the metamodel represents the instances of a Class from the structural metamodel. Its object attributes are defined using the \textit{AttributeLink} class, and associations with other classes are established using the \textit{LinkEnd} class. In BESSER, the object diagram metamodel is primarily utilized for conducting validations or tests on the model. For instance, validating OCL rules over instances of the B-UML model.
    
\textbf{OCL metamodel:} it adds support for defining OCL constraints (e.g. to specify invariants or business rules) on the B-UML models. OCL expressions can be written in plain text and then automatically parsed to create the abstract syntax tree (AST) for expression according to the OCL metamodel \cite{ocl2014object}.
    
\textbf{Graphical user interface metamodel:} 
it focuses on the specification of graphical user interfaces (GUIs) and draws inspiration from the Interaction Flow Modeling Language (IFML) \cite{IFML}, which is widely recognized and adopted in the field of UI modeling. By leveraging IFML, we ensure compatibility and interoperability with existing UI modeling tools and frameworks.
This abstract syntax also enables the definition of GUI components that read and modify elements from a diversity of data sources, including B-UML model elements.

\subsubsection{Concrete syntax}

B-UML models can be created using three types of concrete syntaxes (i.e., notations for the metamodel):
\begin{enumerate}
    \item Textual models conforming to a grammar developed using ANTLR to facilitate the modeling of class diagrams following the syntax of PlantUML\footnote{\url{https://plantuml.com/}}. This also facilitates importing existing PlantUML models into BESSER. Moreover, ChatGPT and other LLMs are familiar with the PlantUML notation and can generate PlantUML models, which is convenient for experimenting in the intersection of modeling and LLMs (see next point).
    \item Models in an image (for example, a photo of a diagram drawn on a whiteboard) can also be processed and transformed into a B-UML model. \newline To achieve this, BESSER employs a LLM to transform the image into a PlantUML model to then generate the B-UML model. A discussion on the quality and accuracy of the recognized models is available here\footnote{\url{https://modeling-languages.com/image-to-uml-with-llm/}}
    \item Using the B-UML python library, a model can be easily created by instantiating the B-UML metaclasses. Several helpers facilitate this task. 
\end{enumerate}

\subsection{Code generators} 
BESSER implements Model-to-Text(M2T) transformations to support code generation. More specifically, it offers a collection of code generators based on the Jinja template engine\footnote{\url{https://palletsprojects.com/p/jinja/}}. These generators read a B-UML model and produce executable code for various technologies and tools such as Django, REST APIs backends or SQLAlchemy for database support. A predefined interface facilitates the addition of new generators.

Listing \ref{lst:jinja_example} shows a glimpse of the Jinja template for generating plain Python classes from B-UML models. At the core of this (and many other) generators you can see a set of nested loops to traverse the different model elements and the properties of every element. 
\begin{lstlisting}[style=python, caption={Example Jinja Template for Python Generation}, label=lst:jinja_example]
{% for class in classes %}
class {{ class.name }}:
    def __init__(self{% for attribute in class.attributes %}, {{ attribute.name }}{% endfor %}):
        {% for attribute in class.attributes %}
        self.{{ attribute.name }} = {{ attribute.name }}
        {% endfor %}
{% endfor %}
\end{lstlisting}

\subsection{Running example}

\begin{figure}
\begin{center}
\includegraphics[width=0.8\textwidth]{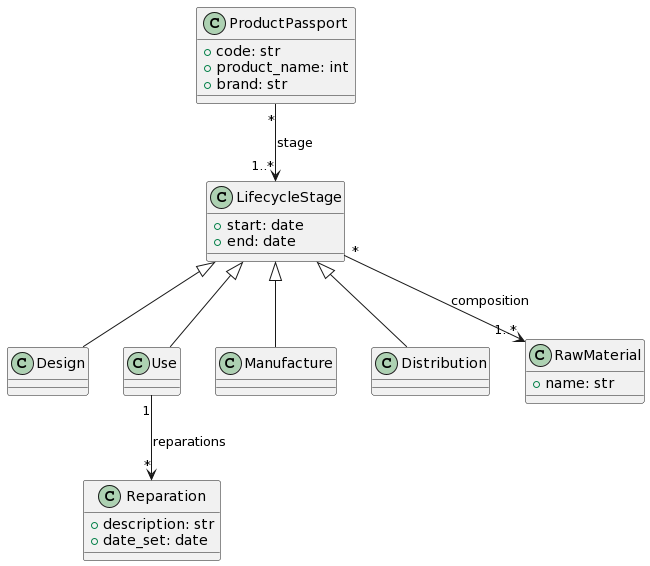}
\caption{DPP model domain} \label{fig:dpp_model}
\end{center}
\end{figure}

As an illustrative case study, we draw inspiration from the Digital Product Passport (DPP) initiative \cite{walden2021digital}, an initiative within the European region aimed at enhancing the circular economy and improving product traceability by collecting data throughout their lifecycle. Illustrated in Figure \ref{fig:dpp_model}, we present a reduced model designed using PlantUML, showcasing key components within this domain, including the \textit{ProductPassport} and its lifecycle stages (\textit{Design}, \textit{Use}, \textit{Manufacture}, etc.). Leveraging the capabilities of BESSER, we utilize the Django code generator to automatically generate a portion of the executable code required for a web application tailored to manage digital passport registrations for products.

\begin{figure}
\includegraphics[width=\textwidth]{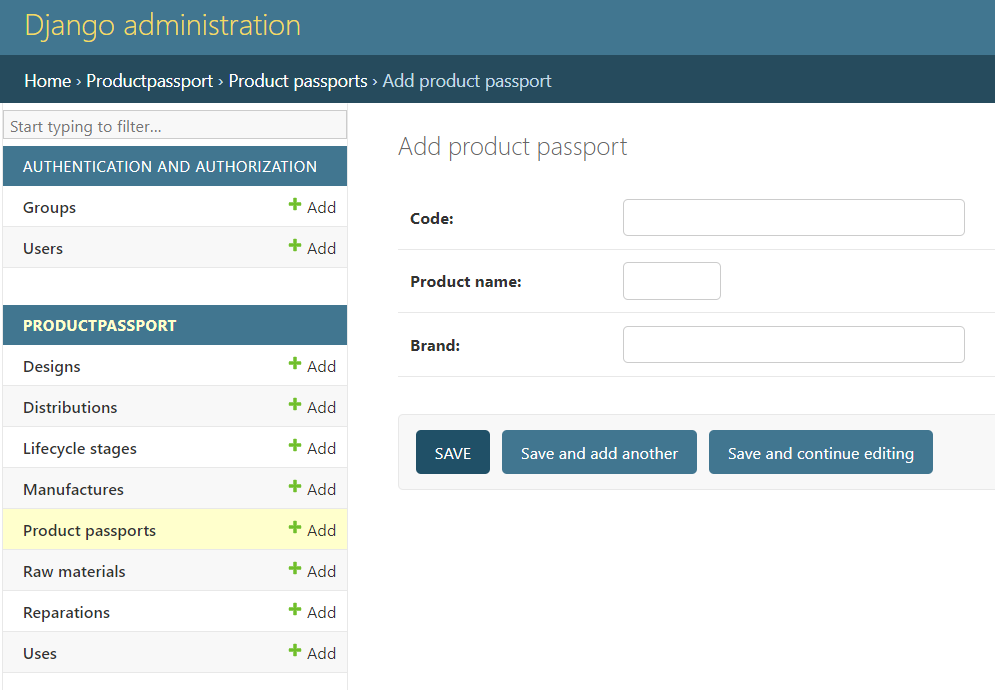}
\caption{Django web application user interface} \label{fig:django}
\end{figure}

The Django code generator uses the DPP domain model as input to generate the model layer code. This code defines model classes that represent the database tables and establishes relationships between them. Figure \ref{fig:django} showcases the user interface of the web application's administration panel that utilizes the generated code. This interface enables the management of database information structured according to the DPP domain model. For instance, the attributes (\textit{code}, \textit{product\_name}, and \textit{brand}) of the \textit{ProductPassport} class are represented as input fields in the creation form in Figure \ref{fig:django}.

The source code and detailed guide of this running example is available in the BESSER examples repository\footnote{https://github.com/BESSER-PEARL/BESSER-examples.git}.

\subsection{Roadmap}
\label{sec:roadmap}
In this section, we briefly describe only a few items, due to space limit, in our roadmap. 

\subsubsection{Modeling smart software}
As discussed in previous sections, we want BESSER to be useful to model smart software systems. This implies being able to model AI components both for the front-end (e.g. chatbots) and the back-end (e.g. recommender systems). While this depends on having a strong core set of modeling components (our core focus at the moment) we have already experimented with the modeling and generation of chatbots (see the next roadmap item) as a concrete first step in this direction.  

\subsubsection{Finite-state machines modeling}
The BESSER Bot Framework (BBF)\footnote{\url{https://github.com/BESSER-PEARL/BESSER-Bot-Framework}} implements finite-state machines (FSM) as a key element in the definition of expressive (chat)bots. In BBF,  bot is defined as a FSM where transitions can be triggered by external events such as matching a user input text with one of the bot`s intents.  Every state has an associated body (a custom Python function) that can run any arbitrary action (e.g., default actions such as replying to the user). As FSMs can be useful in many other modeling scenarios we will ``promote'' the corresponding metaclasses to the main low-code platform. 

\subsubsection{Fully-fledged Web and mobile generators}
While we include a sublanguage for User Interface design and a sample generator for Django, we are aware that to interest a broader user base we should also provide more complete generators, including also generators targeting popular web and mobile development framework such as Flutter or React Native. In the mid-term we would even like to generate a first version (CRUD-like) of the User Interface models from structural models to go from a domain definition to a fully operational mobile application.

\subsubsection{OCL interpreter}
Beyond its current support for the definition of OCL expressions, we want as well to provide an OCL interpreter that can, for instance, be used to validate the constraints on top of scenarios modeled with our object diagram support (kind of what USE does \cite{KuhlmannHG11}). Given the lack of OCL interpreters, especially outside the Java world, we believe this will be a useful contribution. A partial interpreter is already available.

\subsubsection{Flexible modeling}

Traditional definition of domain specific languages (DSL) follows a top-down approach, where both abstract (metamodel) and concrete syntaxes are formulated to specify precisely the model shape and structure. 
However, in some context (e.g., early design phase)
this rigid approach does no longer fit. In order to avoid conflicting situations, the conformance relation that links models and metamodels should be relaxed \cite{hili2017conformance}. 
BESSER will ultimately embed flexibility as a mean to support design changes or natural evolution. However, flexibility should be managed: some part of the model should remain rigid (e.g., for security, regulatory concerns). A mechanism to go from a flexible situation to a rigid one and vice versa will be implemented where, according to a given strategy i) it infers a new metamodel from a given instance or ii) it enforces the conformance and erases all non-conform elements in the model.

\section{BESSER verticals}
\label{sec:verticals}
BESSER provides a core set of components that can be specialized to provide tailored solutions for specific verticals. Depending on the needs of the domain, the specialization can be done via simple subtyping of BESSER concepts or via more complex strategies such as promotional transformations \cite{Jacome-Guerrero20}.

As an example, we are developing \textit{BESSER for Clima} as part of the Climaborough\footnote{\url{https://climaborough.eu/}} EU project. The goal  by providing the necessary tools to produce a dashboard showcasing the performance of climate solutions and the progress on reaching various climate related goals such as net neutrality. Beyond using BUML and the SQL generator to produce the necessary database schema, a visualisation generator is being developed. This generator will consume models that associate data with visualisations and produce visualisations that can be directly integrated into JavaScript pages. The usage of JavaScript libraries covers a large range of visualisation possibilities and is supported by the state-of-the-art web development frameworks.

Additional verticals (e.g. for digital twin architectures, for digital product passports in a lifecycle context, etc) will be developed as well.

\section{Discussion}
\label{sec:discussion}
This section briefly comments on three design decisions that influence the development of BESSER.

\subsubsection{Python over Java}
Most modeling tools are built in Java, and especially around the Eclipse Modeling Framework \footnote{\url{https://eclipse.dev/modeling/emf/}} ecosystem. Nevertheless, given that Python is the default language for all types of Machine Learning (ML) libraries and frameworks we decided to opt for Python as core language for BESSER. This facilitates the integration of ML components in it.

\subsubsection{Internal DSLs over external ones}
Based on our experience building other modeling tools \cite{DanielC24}, we decided that BESSER would be better suited to be defined as an Internal DSL, i.e. a language implemented on top of an underlying programming language  \cite{ghosh2011dsl}, instead of building it as a completely separated language. The key reason is that we want BESSER to cover as many scenarios as possible and this would often imply relying on general programming langauge, e.g. to specify complex behaviour. Defining BESSER as an internal DSL facilitates this. Similarly, we opted to choose other common language processing tools (such as Jinja and ANTLR) to implement the generators and parsers. As an additional advantage, they have both a strong and active community behind, which is important for the long-term sustainability and evolution of BESSER.

\subsubsection{Why not just fully UML compliant?}
As discussed before, B-UML is inspired but not a clone of UML. This gives us the freedom to keep the parts we need but disregard those that we do not. It also allow us to integrate new languages that cover aspects that UML does not support well (e.g. UI  modeling) or add new concepts or properties that are needed when modeling in the context of low-code development. 

\section{Conclusion}
\label{sec:conclusion}
This paper presents BESSER, a new open source low-code platform targeting not only traditional software systems but also smart components. We have also presented a short-term roadmap and a discussion on the design decisions that, so far, have shaped the architecture of the tool. We hope the modeling community finds our platform useful and decides to adopt it (and even collaborate to its development) in future projects and initiatives around the world of modeling, low-code and, in general, model-driven engineering approaches.

As longer-term future work, we plan to add smart components to BESSER itself, facilitating the creation of B-UML models via low-modeling techniques \cite{cabot2024lowmodeling} such as chatbots or automatic model inference from (semi)structured data.

%
%

\bibliographystyle{splncs04}

%
\end{document}